\documentclass[twocolumn]{revtex4}  

\usepackage{graphicx}  

\begin{document}

\title{GaAs photonic crystal cavity with ultra-high Q: microwatt nonlinearity at 1.55 $\mu $m}

\author{Sylvain Combri\'{e}}
\email{sylvain.combrie@thalesgroup.com}
\author{Alfredo De Rossi}
\author{Quynh Vy Tran}
\affiliation{Thales Research and Technology, route d\'{e}partementale 128, 91767 Palaiseau, France}%

\author{Henri Benisty}
\affiliation{Laboratoire Charles Fabry, CNRS, Institut d'Optique Graduate School, 91127 Palaiseau, France}

\date{\today}%

\begin{abstract}
We have realized and measured a GaAs nanocavity in a slab photonic crystal based on the design by Kuramochi et al. [Appl. Phys.Lett., \textbf{88} , 041112, (2006)]. We measure a quality factor Q=700,000, which proves that  ultra-high Q nanocavities are also feasible in GaAs. We show that, due to larger two-photon absorption (TPA) in GaAs,  nonlinearities appear at the microwatt-level and will be more functional in gallium arsenide than in silicon nanocavities.
\copyright{Optical Society of America}
\end{abstract}

\maketitle

\indent The achievement of quality factors of Q $\approx10^6$ \cite{Kuramochi1,Noda2} in micron-sized nanocavities carved in two-dimensional photonic crystals (PhC) opens perspectives for all-optical signal processing. The unique property of PhC cavities is their ultra small volume (on the order of 0.1 $\mu$ $m^3$) which, combined with a high Q factor, dramatically enhances light-matter interaction. This is highly desirable for fundamental investigations in cavity quantum electrodynamics (QED) in condensed matter, where a single emitter, e.g. a quantum dot (QD), is strongly-coupled to a single optical mode. As emission linewidth of a single QD can be as small as a few $\mu$eV at low temperature\cite{Cassabois}, reaching a comparable linewidth for the cavity resonance is clearly desirable. Besides QED and applications to micro-lasers and non-conventional laser sources such as single photon emitters for quantum key distribution, the strong enhancement of light-matter interaction is fundamental in order to miniaturize a broad class of devices such as optical sensors and optical modulators. Furthermore, high Q-values are compatible with broadband operation when a so-called CROW type waveguide is built from multiple identical cavities \cite{NotomiNature2007}. For slow light purposes or for nonlinearity enhancement, the intrinsic Q factor determines the maximum amount of delay/enhancement which can be achieved.The state-of-the-art in high-Q PhC resonators has been achieved by NTT and Kyoto's teams \cite{Kuramochi1,Noda2} in silicon-based structures.
There is a widespread consensus that silicon processing technology is superior to III-V semiconductor and therefore that there is little chance that PhCs based on III-Vs approach the state-of-the-art. On the other hand, III-Vs offer unique opportunities in photonics, in particular emission/amplification of light.
\newline\indent In this letter, we show that GaAs can reach Q values similar to those of silicon, namely Q$>$700,000. 
This result opens new perspectives for realisations combining the features of III-V materials with the attractive properties of PhC, including ultra-low-power nonlinear optics. Microwatt-level nonlinear operation can therefore be envisioned from the $\approx5 \mu$W threshold power value obtained in our previous results (\cite{Weidner4,Combrie6}) for the lower quality factor Q$\approx$246,000. With the linear bandwidth set at $\approx$1 GHz by the cavity, nonlinear processing of microwatt optical signal in the 1 MHz - 1 GHz window can be achieved. From the scaling laws of the various effects (Kerr, free carrier plasma, two-photon absorption, thermo-optic), we pinpoint the more favorable capabilities of GaAs in this respect.

\begin{figure}[h]
\includegraphics[width=8.5cm]{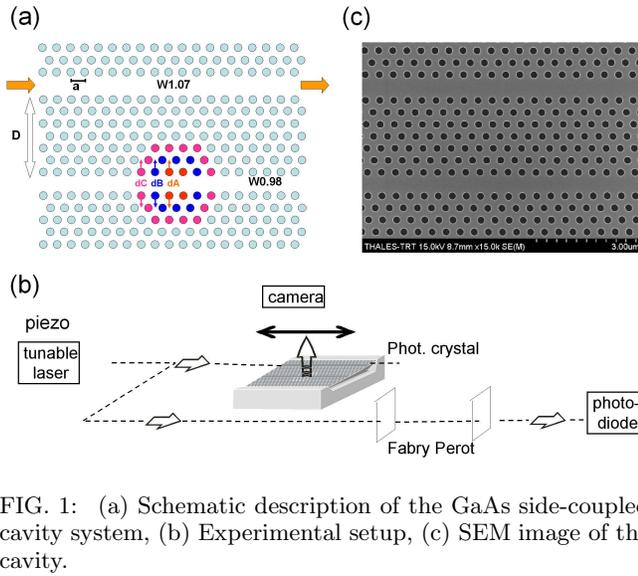}
\caption{ (a) Schematic description of the GaAs side-coupled cavity system, (b) Experimental setup, (c) SEM image of the cavity.}
\label{fig:design}
\end{figure}

Our strategy was to consider the design that ensured the highest theoretical Q factor and which has also been implemented successfully in Si structures. At the time we designed the cavity, this was the design proposed in ref. \cite{Kuramochi1}. Our underlying idea is that this structure must be also the most robust against fabrication tolerances. Three dimensional finite-difference-time domain (FDTD3D) modeling predicted a Q factor of about 100 millions.
We used a 186-nm-thick GaAs membrane, and a basic lattice pitch a=420 nm. As shown in Fig.\ref{fig:design}a, the access waveguide is designed as W1.07 (W1 refers to the single missing row waveguide along the $\Gamma$K direction of the photonic crystal) while the width of structure supporting the cavity is \textit{Wx}, with \textit{x} variable but $x<1$. The hole shifts defining the cavity are 9 nm, 6 nm and 3 nm. The waveguide-cavity separation varies between 7 and 9 rows.

\begin{figure}[h]
\includegraphics[width=8.5cm]{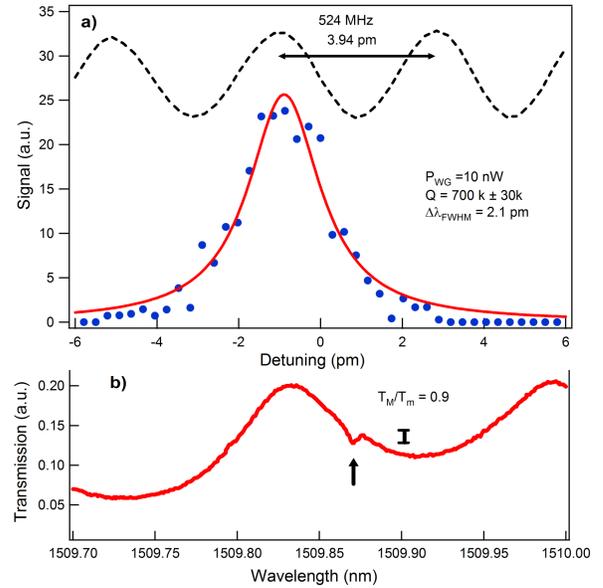}
\caption{Spectra of the resonance of the cavity n$^{\circ}$5. The power coupled into the waveguide is about 10 nW.  a) Detected signal from the camera  as a function of the detuning (markers) and Lorentzian fit (solid line). Transmission of the Fabry P\'{e}rot interferometer used as reference (dashed line). b) Transmitted signal through the waveguide. The cavity resonance is pointed by the arrow.}
\label{fig:exp_setup_sample}
\end{figure}

\indent We used a compact and efficient 100kV e-beam writer nB3 (NanoBeam Ltd., Cambridge, UK) to define the patterns in the top resist layer, the rest being unchanged. The good results obtained validate the qualities of this tool. Inductively-Coupled Reactive-Ion-Etching \cite{Combrie5} was used to perform GaAs/GaInP vertical etching.
As for measurements, we used a tunable laser source (Tunics from Nettest) operated in the fine scanning mode by applying an external voltage. The wavelength shift is monitored with a low-finesse Fabry-Perot (FP) interferometer (Fig. \ref{fig:design}b) with 28.6-cm-spaced mirrors (free-spectral range is 526 MHz). 
We have fabricated nine cavities with slightly modified parameters (controlling the coupling strength to the waveguide) and measured 4 of them (n$^{\circ}$ 3, 4, 5 and 7). Cavities n$^{\circ}$5 and n$^{\circ}$7 have been designed for maximizing Q with purposely weak coupling (x=0.98 and spacing is 9 rows), the two other cavities (n$^{\circ}$3 and n$^{\circ}$4) have non optimal parameters and were designed with a stronger waveguide to cavity coupling. Measurements have been performed at different power levels, in order to identify the linear regime whereby the Q factor saturates as power is further reduced. Each of these measurements has been repeated 10 times. The uncertainty is deduced from the fitting procedure and also corresponds to the fluctuation across measurements.
\indent Figure \ref{fig:exp_setup_sample} shows the measurement for the cavity n$^{\circ}$5. The signal detected from the top of the cavity provides a peak at resonance (Fig. \ref{fig:exp_setup_sample}a), while the transmitted waveguide signal displays a corresponding dip (Fig. \ref{fig:exp_setup_sample}b) whose depth is indicative of coupling conditions\cite{Combrie6}. A largely sub-critical coupling was observed (the  minimum/max. transmission ratio at resonance being $T_m/T_M\approx90\%$). The value for the loaded Q factor (Lorentzian fit, averaged over several measurements, Fig. \ref{fig:exp_setup_sample}a) was $Q$=700,000$\pm$30,000 ($\Delta\nu$=280 MHz, $\Delta\lambda$=2.1pm). The estimated optical power coupled into the waveguide is here about 10 nW. The fraction of power $P_c/P_{wg}$ actually coupled into the cavity is $P_c/P_{wg}=2(\sqrt{T_m/T_M}-T_m/T_M)\approx0.1$, which makes 1 nW. We had to stick to such small values in order to prevent broadening induced by nonlinear absorption. We have also measured Q$\approx$ 700,000 in cavity n$^{\circ}$7. 
In spite of the progress made in PhC microcavities, the gap between theoretical expectation and experimental measurements of the Q factor is large, here is almost two order of magnitude. This holds also for silicon.
This suggests that the limit of processing capability is close, even for state-of-the-art e-beam systems. The important point of this work is that the handicap of III-V processing with respect to silicon has almost vanished. We are convinced that the residual gap separating us from the world record is rather related to residual structural disorder (e-beam lithography and plasma etching). We think that other factors, such as residual absorption in the material, are still negligible in this spectral range. We also believe that our process (high density plasma) generates a relatively low number of surface defects so that an additional source of losses is avoided.

\begin{figure}[h]
\includegraphics[width=8.5cm]{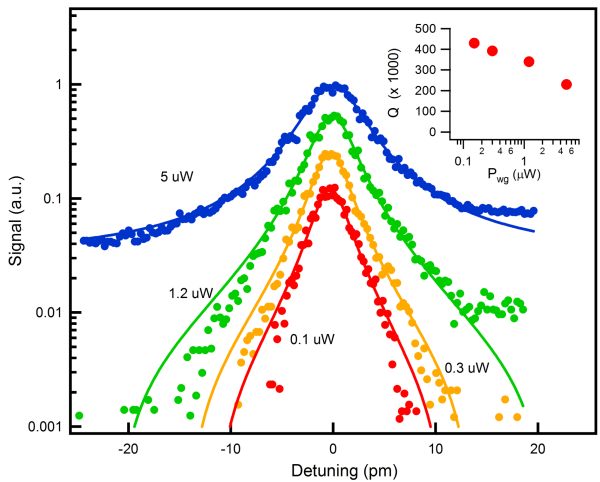}
\caption{Cavity n$^{\circ}$3. Signal detected from the cavity (log. scale) as a function of the detuning at different optical power coupled in the waveguide ($P_{wg}$). The curves are shifted vertically for clarity. Inset: dependence of the Q factor on $P_{wg}$.}
\label{fig:nonlinear}
\end{figure}

A Q scaling from our previous work \cite{Weidner4} indicates that for such high values of the Q factor, we enter in a nonlinear regime at the microwatt level in terms of power flow in the waveguide. This is confirmed in Fig. \ref{fig:nonlinear}, where we report the dependence of the lineshape of another cavity than that of Fig.\ref{fig:exp_setup_sample} as a function of the power coupled in the waveguide. The onset of nonlinearity (here Two Photon Absorption) appears as the power coupled to the waveguide is on the order of 0.3$\mu$W.
However, as underlined by this work and previous ones \cite{Uesugi7}, both thermal and electronic nonlinearities are involved: Kerr effect, two photon absorption (TPA) and the resulting index changes due to free-carrier plasma and thermo-optic effect.
For thermo-optic effects, at first sight, silicon is a better heat conductor. We show below that this appearance is largely offset by the intrinsic lower operation point of GaAs, giving to this latter material a clear niche for ultra-low power optical manipulations. Let us detail how the nonlinear operation can be performed to further evaluate GaAs vs. Si. 
Firstly, in transient operation, the thermo-optic effect has a time-dependent spatial extent \textit{x}, given by the thermal diffusivity  $C_{p}$, obeying the $x\approx\left(C_{p}\tau\right)^{1/2}$ scaling (where $C_{p}$ is 0.31 and 0.78 $m^2/s$ for GaAs and Si respectively). In the MHz-GHz range \textit{x} is submicrometric and the heat resulting from TPA will still reside within the cavity, not even spreading beyond one PhC lattice pitch\cite{derossi2008}. This implies the virtual impossibility to implement an extra thermal sink at these high Q/low nonlinear threshold values. 
Secondly, the TPA damping threshold (at which the induced damping halves Q) is evaluated in terms of power coupled through the cavity $P_c$ as: $P^{th}_c$=$(4\pi^{2}V_{TPA})/(\lambda^{2}Q^2\beta)$ $\propto\beta^{-1}$  Here the TPA constant is $\beta$ = 10 cm/GW for GaAs and the nonlinear effective volume is about 0.4 $\mu m^3$. This amounts to approximately 100 nW for a mean quality factor Q = $7\:10^{5}$. This is in quite good agreement with data measured on cavity n$^{\circ}$3 in Fig. \ref{fig:nonlinear} where the estimated power at which the Q-factor drops is $P^{th}_{WG}\approx2\:\mu W$ in the waveguide and therefore much lower in the cavity, due to the weak coupling. We have also measured the power for cavity n$^{\circ}$5 and found that Q drops to 500,000 as power in the waveguide is raised to $1\:\mu W$, i.e. 100 nW in the cavity.
Thirdly, the impact of thermo-optic effects derives from the index shift  $\Delta$n which is proportional to the thermo-optical coefficient $n_T$, the thermal resistance and the amount of power absorbed, which turns out to be governed by the ratio $n_T/(\beta\:C_p)$.
 This is a key point: obviously, a stronger TPA coefficient and a lower power threshold weakens the thermal burden. Of particular interest here is the fact that the $\approx$4 times weaker $n_{T}/C_{p}$ ratio of silicon is well offset by the $>$10 times larger TPA coefficient $\beta$ of GaAs.

Fourthly, the carriers generated by TPA induce a negative index shift counteracting the Kerr effect. This is governed by the carrier recombination time $\tau_{rec}$, which can be considered to be fast ($\tau_{rec}$ is in 10-100 ps time scale\cite{Bristow11,Uesugi7}). The amount of energy stored in the cavity \textit{W} at which the plasma-induced index change takes over the Kerr effect is $W_{th}\approx{n_{2}m^{*}}/(\beta\tau_{rec})$, with $m^*$ the effective electron mass and $n_2$ the Kerr coefficient. With the values 1.6 and 0.45 $cm^2/GW$ for $n_{2}$ in GaAs and Si respectively, the crossover energy ratio is: $W_{th,Si}/W_{th,GaAs}\approx$15$\times\tau_{rec,GaAs}/\tau_{rec,Si}$. Moreover, the net ratio is almost entirely compensated when considering typical values for $\tau_{rec}$: 100 ps for Si and 10 ps for GaAs\cite{Bristow11}, resulting in similar crossover powers. Usually, Kerr effect is sought for optical manipulation and TPA seen as an hindrance. Taking an opposite approach, i.e. exploiting nonlinear cavity damping, it becomes advantageous to use GaAs : as it operates at a much lower power, it features less index shift from the Kerr effect than silicon.

\indent In conclusion, we showed that an ultra-high Q nanocavity akin to those elaborated in silicon is also feasible in GaAs, making GaAs devices with Q$\approx10^6$ fully plausible. Additionally, we stressed the possibility to operate at the microwatt level for nonlinear operation, through nonlinear damping based on two-photon-absorption. 
Importantly, we substantiated the fact that thermal effects inflict less severe penalties when operating nanocavities based on GaAs as compared to those based on Si.
\newline\indent We acknowledge the support of the SESAME action of Conseil G\'{e}n\'{e}ral Ile de France for key equipments used in this work.
The TRT staff ackowledges the financial support of the European Commission through the IST Project "QPhoton".

\end{document}